%Paper: astro-ph/9309056
%From: "T. Buchert" <TOB@IBMA.ipp-garching.mpg.de>
%Date: Thu, 30 Sep 93 12:58:32 MET

\magnification=1200
\tolerance=1400
\overfullrule=0pt
\baselineskip=15pt
\font\mfst=cmr9
\font\mfs=cmr9 scaled \magstep 1
\font\rma=cmr9 scaled \magstep 2
\font\rmm=cmr9 scaled \magstep 3
\def\ueber#1#2{{\setbox0=\hbox{$#1$}%
  \setbox1=\hbox to\wd0{\hss$\scriptscriptstyle #2$\hss}%
  \offinterlineskip
  \vbox{\box1\kern0.4mm\box0}}{}}

\def\hn{$\;{\rm h}^{-1}$}

\def\R{\rm I\kern-.18em R}
\def\etal{{\it et al. }}
\topskip 6 true cm
\pageno=0

\centerline{\rmm TESTING HIGHER-ORDER LAGRANGIAN}
\smallskip
\centerline{\rmm PERTURBATION THEORY}
\smallskip
\centerline{\rmm AGAINST NUMERICAL SIMULATIONS -}
\bigskip
\centerline{\rmm 1. PANCAKE MODELS}
\bigskip\bigskip
\centerline{\rmm by}
\bigskip\bigskip
\centerline{\rmm T. Buchert$^{1}$, A.L. Melott$^{2}$, A.G.
Wei{\ss}$^{1}$}
\vskip 1.5 true cm
\centerline{\rma $^{1}$Max-Planck-Institut f{\"u}r Astrophysik}
\smallskip
\centerline{\rma Karl-Schwarzschild-Str. 1}
\smallskip
\centerline{\rma 85740 Garching, Munich}
\smallskip
\centerline{\rma F. R. G.}
\bigskip
\centerline{\rma $^{2}$Department of Physics and Astronomy}
\smallskip
\centerline{\rma University of Kansas}
\smallskip
\centerline{\rma Lawrence, Kansas 66045}
\smallskip
\centerline{\rma U. S. A.}

\vskip 1 true cm
\centerline{\mfs submitted to {\it Astron. Astrophys., Main
Journal}}
\vfill
\eject

\topskip= 2 true cm
\centerline{\rmm Testing higher-order Lagrangian perturbation theory}
\smallskip

\centerline{\rmm against numerical simulations -}

\medskip

\centerline{\rmm 1. Pancake models}
\bigskip
\centerline{\rma by}
\bigskip
\centerline{\rma T. Buchert, A.L. Melott, A.G. Wei{\ss}}
\vskip 0.8 true cm
\noindent
{\narrower{\mfst
{\baselineskip=13pt
{\mfs Summary:}
We present results showing an improvement of the accuracy of
perturbation
theory as applied to cosmological structure formation for a useful range
of
quasilinear scales.
The Lagrangian theory of gravitational instability of
an Einstein-de Sitter dust cosmogony
investigated and solved
up to the third order in the series of papers by
Buchert (1989, 1992, 1993a), Buchert \& Ehlers (1993), Buchert (1993b),
Ehlers \& Buchert (1993),
is compared with numerical simulations.
In this paper we study the dynamics of pancake
models as a first step.
In previous work (Coles \etal 1993, Melott \etal 1993, Melott 1993)
the accuracy of
several analytical approximations for the modeling of large-scale
structure in the mildly non-linear regime was analyzed in the same way,
allowing for direct comparison of the accuracy
of various approximations. In
particular, the ``Zel'dovich approximation''
(Zel'dovich 1970, 1973, hereafter ZA) as a subclass of
the first-order Lagrangian perturbation solutions was found to
provide an excellent approximation to the density field in the mildly
non-linear regime (i.e. up to a linear r.m.s. density contrast of
$\sigma \approx 2$). The
performance of ZA in hierarchical clustering models
can be greatly improved by truncating the initial power spectrum
(smoothing the initial data).
We here explore whether this approximation can be further
improved with higher-order corrections in the displacement mapping
from homogeneity.
We study a single pancake model (truncated power-spectrum with
power-index $n=-1$) using cross-correlation statistics employed in
previous work.

\noindent
We found that for all statistical methods used the higher-order
corrections improve the results obtained for
the first-order solution up to the stage when $\sigma$(linear theory)
$\approx 1$.
While this improvement can be seen for all spatial
scales, later stages retain this feature only above a certain scale
which is increasing with time.
However, third--order is not much improvement over
second--order at any stage.
The total breakdown of the perturbation approach is observed at
the stage, where $\sigma$(linear theory)$\approx 2$, which corresponds
to the onset of hierarchical clustering. This success is found at a
considerably higher
non-linearity than is usual for perturbation theory.
Whether a truncation of
the initial power-spectrum in hierarchical models retains this
improvement will be analyzed in a forthcoming work.

}}}

\vfill\eject
\topskip= 0 true cm

\noindent{\rmm 1. Introduction}
\bigskip\bigskip
\noindent
In most recent research in cosmology, the gravitational
instability picture of the standard Friedman-Lema\^\i tre cosmogonies
has been used to describe the earliest stages of the formation of
inhomogeneities in the Universe (see, e.g, Peebles 1980 and ref.
therein). The first quantitative
exploration used linear perturbation theory to
make predictions from theory.
Although there are serious restrictions of applicability of this
perturbation approach (e.g., the density contrast $\delta =
{\rho - \rho_H \over \rho_H}$, where $\rho_H$ is the homogeneous
density of the background cosmogony, should be much less than $1$),
the solutions are still applied to normalize inhomogeneous
cosmological models,
i.e., to determine the amplitude of the initial fluctuation field
in comparison with the density contrast we observe today.
We now know that first-order solutions of this perturbation
approach cannot be used as an approximation of the present day
density field.
Only, if we smooth the field with an extremely large smoothing length
of the order of $100$\hn Mpc (Coles \etal 1993), a comparison with
non-linearly evolved density fields can be meaningful.
This perturbation approach is Eulerian, i.e., the perturbations are
evaluated as functions of Eulerian coordinates.

\medskip
In the $70$'s Zel'dovich realized the limitations of this
approach and proposed an extrapolation of the Eulerian linear
solutions into the mildly non-linear regime by employing the
Lagrangian picture of continuum mechanics (Zel'dovich 1970, 1973).
In his model the motion of the continuum is similar to the purely
kinematical motion of a system of particles which move under inertia.
The model can be mapped to this system by a suitable transformation of
dependent and independent variables (Shandarin \& Zel'dovich 1984,
1989).
The success of Zel'dovich's extrapolation for the description of
typical
patterns of the large-scale structure has been widely recognized,
although not
generally understood.
It can be traced back to the fact that the Lagrangian description is
based on following the trajectories of fluid elements, while the density
itself does not appear as a dynamical variable in this picture and
can be integrated exactly along the trajectories of a given
solution.
Consequently, the density is not limited in a Lagrangian perturbation
approach, only the gradient of the displacement field has to be small
compared with the expansion factor of the homogeneous background
(Buchert 1992, Ehlers \& Buchert 1993).
Moreover, Zel'dovich's model can be
systematically derived by solving the Lagrangian evolution equations to
the first-order (Buchert 1992). It also provides a class of exact
three-dimensional solutions in the case where the collapse is
maximally anisotropic (i.e., in the case where locally
two of three eigenvalues
of the fluid's deformation tensor vanish) (Buchert 1989).
The plane-symmetric case is contained in a subclass of these
solutions; first-order Lagrangian perturbation solutions
are the general solutions of the full system in this case.

\medskip
Lagrangian perturbation solutions are now available
for various backgrounds and up to the third order in the deviations
from homogeneity (see Buchert 1993b and references therein).
Parallel efforts to investigate Lagrangian perturbation solutions
are made by Moutarde \etal (1991), Bouchet \etal (1992), Lachi\`eze-Rey
(1993), Gramann (1993), Giavalisco \etal (1993), Croudace \etal (1993),
see also Bertschinger \& Jain (1993).

\medskip
Coles, Melott and Shandarin (1993) conducted a series
of tests of analytical approximations
by cross-correlating them with N-body
simulations. They tested the
Eulerian linear perturbation solution and the
``Zel'dovich approximation'' (hereafter ZA)
as a subclass of the irrotational
Lagrangian linear solution. They found that the latter was
considerably more successful than the former. Considerable further
improvement is made if the ZA is applied
not to the full power spectrum on the whole range of spatial scales, but
only to a truncated body of the spectrum. The truncation removes
unwanted non-linearities, which are evolved beyond the range of
applicability of the model
(Melott and Shandarin 1990, Beacom
\etal 1991, Kofman \etal 1992, Coles
\etal 1993).
The shape of high-frequency filters imposed on the power-spectrum was
found to be important (Melott \etal 1993).
Most successful in respect of improvement of the cross-correlation
statistics was the use of Gaussian filters,
but if one wishes to work backwards
from evolved to initial state,
a sharp truncation (step function) in $k$ is
preferable (Melott 1983).

\medskip
Since closer study
of analytical approximations reveals both limitations and
powerful properties, it is worthwhile to study generalizations of
Zel'dovich's mapping to improve the cross-correlation with numerical
simulations. In this way the modeling of large-scale structure
is effective and can be used to provide initial and boundary conditions
for the evolution of small-scale structure within the large-scale
environment, which could be modelled, e.g., by fully hydrodynamical
simulations.
Although the applicability of the analytical models under
study
is not limited to a particular type of spectrum (the ``Zel'dovich
approximation'' was originally used for modeling the standard ``pancake
picture'', see Shandarin \& Zel'dovich 1989), we here do concentrate
our comparison on ``pancake models'', i.e., models which do not
involve small-scale power in the initial data. We do this
as a first step to properly understand the limitations of higher-order
corrections to Zel'dovich's mapping. Also, pancake models can be
understood as generic archetype of hierarchical models which involve
pancaking on all spatial scales (Kofman \etal 1992). Because the models
we
study here already have a truncated initial power spectrum, we do not
apply
additional truncation as in Coles \etal (1993) or Melott, Pellman and
Shandarin
(1993). In a second step,
we shall later analyze various power spectra which are evolved deeply
into the
non-linear regime in order to probe the performance of the Lagrangian
approximations in the case of hierarchical models (paper II
{\sl in
preparation}).
\medskip
Further work on other approximations is also in preparation: the
frozen--flow
approximation (Matarrese \etal 1992) and the adhesion approximation
(Kofman
 \etal
1992 and references therein) will be given a similar treatment.

\medskip
Another advantage of analytical models is the possibility of
high-spatial re\-so\-lution of the density field. Analytical mappings
of a continuum offer in principle infinite resolution in contrast to
numerical N-body simulations. We shall explain that
models with sufficiently smooth initial data like pancake models
can be resolved easily to, e.g., an effective resolution of
$2048^3$ particles. However, as we shall see later, the accuracy of this
effort
would be limited to early stages of non-linearity.

\bigskip\bigskip\bigskip

\noindent{\rmm 2. Numerical realization of pancake models}
\bigskip\medskip
\noindent
We specify initial data in terms of a power spectrum
${\cal P}(k)$ (as a function of
comoving wavenumber $k=\vert \vec k \vert$)
of Gaussian density perturbations of the form:
$$
{\cal P} (k)=
<\vert \delta_{\vec k} \vert^2 >
\propto \vert \vec k \vert^n\;\;,\eqno(1)
$$
where $\delta_{\vec k}$
is the discrete spatial Fourier transform of the density
contrast $\delta$. We shall take $n=-1$, and truncate the power spectrum
at a characteristic frequency $k_c$. In this way we will probe
the epoch of first pancaking.
We also define the non-linearity scale
$k_{nl}(t)$:
$$
a^2(t) \; \int_0^{k_{nl}(t)} \; d^3 k \; {\cal P}(k) \;=\;1\;\;,\eqno(2)
$$
where $k_{nl}(t)$ is decreasing with time as successively larger scales
enter the non-linear regime; $a(t)$ is the scale factor of the
homogeneous background: for the initial data we set $k_c=4k_f$, and
$a(t_i):=1$ where $k_f$ is the fundamental mode of the simulation box.
We emphasize that we give initial data early (at the r.m.s.
density contrast of $\sigma_i (k_c)= 0.01$)
to guarantee an objective modeling of the collapse
of first pancakes in the model. One of us (Melott 1985, 1987) has
emphasized the importance of beginning numerical simulations
at a highly linear stage to avoid an artificial delay of the collapse
time of first objects. This delay is inherent in Zel'dovich's
displacement
mapping which is used to initialize the numerical simulation
(compare Blanchard \etal (1993) for a discussion of this problem).
{}From several first studies of higher-order perturbation solutions
we know that the collapse
is significantly accelerated by the higher-order corrections
even in a generic field (Buchert \etal 1993).
An objective test of this feature can only be conducted by giving
the initial data early. The results of making
too late a start with too
high an amplitude are not detectable in the two--point correlation
function since this
is mostly determined by long waves just going non-linear
(Beacom \etal 1991;
Melott \& Shandarin 1993).

\medskip
The evolution of inhomogeneities is modelled in a flat
Einstein-de Sitter background ($a(t)=({t \over t_i})^{2/3}$).
We evolve the field with an enhanced PM (particle-mesh) method
(Melott 1986) using
$128^3$ particles
each on a comoving $128^3$ mesh with periodic boundary conditions.
This method makes the code resolution-equivalent to a traditional
PM code with $128^3$ particles on a $256^3$ mesh (see Weinberg \etal
(1993)).

\medskip
In what follows we will specify the stages
we study by the {\it Eulerian linear theory}
amplitude which is just proportional to the scale factor in these
$\Omega=1$
models. (The r.m.s. density contrast $\sigma_2$ denotes
$\sigma$(linear theory) in the numerical simulation throughout this
paper.)
Of course, the real density contrast is larger due to mode coupling
effects.
\medskip
{}From this simulation we extract six stages, the first at the
epoch when pancake formation is just beginning ($\sigma_2=0.25$), the
second at the moment when the first pancakes have collapsed
($\sigma_2=0.5$), the third at a stage when
the cellular connection of different pancakes is established
(corresponding to the contemporary epoch according to the standard
normalization)
($\sigma_2=1$), the fourth and fifth at more
advanced
stages, where secondary shell-crossings have developed ($\sigma_2=1.2$
and $\sigma_2=1.5$),
and the last at a stage when the
non-linearity scale has dropped to $k_{nl}=2k_f$ ($\sigma_2 = 2$).
This stage corresponds to the onset of hierarchical clustering
due to merging of pancakes.

\bigskip\bigskip\bigskip

\noindent{\rmm 3. The Lagrangian perturbation solutions}
\bigskip\medskip
\noindent
In what follows we make use of a restriction of the
initial data commonly used in realizations of large-scale
structure models as well as in the numerical realization quoted above.
We require that, initially, the peculiar-velocity $\vec u(\vec
X,t_i)$ be proportional to the peculiar-acceleration
$\vec w(\vec X,t_i)$:
$$
\vec u (\vec X,t_i) = \vec w (\vec X,t_i) t_i \;\;\;,\eqno(3)
$$
where we
have defined the fields as usual (compare Peebles 1980, Buchert 1992).
This restriction has proved to be appropriate for the purpose of
modeling large-scale structure since, for irrotational
flows, the peculiar-velocity field tends to be parallel to the
gravitational peculiar-acceleration after some time.
The reason for this tendency is related to the existence of growing
and decaying perturbations in the linear regime,
the growing part supports the
tendency to parallelity.
The assumption of irrotationality should be adequate down to the
non-linearity scale.

\medskip
Henceforth, a comma denotes derivative with respect to
Lagrangian coordinates, and
$\Delta_0:=\nabla_0^2$, where
$\nabla_0$ denotes
the nabla operator with respect to Lagrangian coordinates.
\noindent
We shall use in the following the initial peculiar-velocity potential
${\cal S}$ defined as $\vec u (\vec X,t_i) =: \nabla_0 {\cal S}(\vec
X)$. The initial peculiar-gravitational potential $\phi$,
$\vec w (\vec X,t_i) =:-\nabla_0 \phi (\vec X)$ is related to it as
${\cal S} = - \phi t_i$ (eq. (3)).
Further, we introduce the symbols
$I({\cal S}_{,i,k}) = tr({\cal S}_{,i,k}) = \Delta_0 {\cal S}$,
$II({\cal S}_{,i,k}) = {1 \over 2} \lbrack(tr({\cal S}_{,i,k}))^2 -
tr(({\cal S}_{,i,k})^2)\rbrack$
and $III({\cal S}_{,i,k})=\det({\cal S}_{,i,k})$ which
denote the three principal scalar invariants of the tensor
$({\cal S}_{,i,k})$.

\noindent
We introduce the field of trajectories
$\vec x = \vec f (\vec X,t)$, where $\vec x$ denote Eulerian
and $\vec X$ Lagrangian coordinates. We express the solutions
in a comoving reference system $\vec q = \vec F (\vec X,t)$, where
$\vec q = \vec x / a(t)$ denote comoving Eulerian coordinates.

\medskip
With a superposition ansatz for Lagrangian perturbations of an
Einstein-de Sitter background
the following family of trajectories $\vec q = \vec F
(\vec X,a)$ as irrotational
solution of the Euler-Poisson system up to the third
order in the perturbations from homogeneity has been given by one of us
(Buchert 1993b).
The general set of
initial data $(\phi (\vec X), {\cal S} (\vec X))$ is
only restricted according to ${\cal S} = - \phi t_i$ (eq. (3)):
$$
\vec F = \vec X \;+\;
q_1 (a) \; \nabla_0 {\cal S}^{(1)} (\vec X) \;+\; q_{2}
(a) \; \nabla_0 {\cal S}^{(2)} (\vec X)
$$
$$
+\; q_{3}^{a} (a) \;
\nabla_0 {\cal S}^{(3a)} (\vec X) \;+\; q_{3}^{b} (a) \;
\nabla_0 {\cal S}^{(3b)} (\vec X) \;-\;
q_{3}^{c} (a) \; \nabla_0 \times {\vec {\cal S}}^{(3c)} (\vec X)
\;\;,\eqno(4)
$$
with the time-dependent coefficients expressed in terms of the
expansion function $a(t) = ({t \over t_i})^{2/3}$:
$$
\eqalignno{
q_1 &= \left({3 \over 2}\right) (a - 1) \;\;\;, &(4a) \cr
q_{2} &= \left({3 \over 2}\right)^2
(-{3 \over 14} a^2 + {3 \over 5} a - {1 \over 2}
+ {4 \over 35} a^{-{3 \over 2}}) \;\;\;, &(4b) \cr
q_{3}^{a} &= \left({3 \over 2}\right)^3
(-{1 \over 9} a^3 + {3 \over 7} a^2 - {3 \over 5} a
+ {1 \over 3} - {16 \over 315} a^{-{3 \over 2}}) \;\;\;,&(4c) \cr
q_{3}^{b} &= \left({3 \over 2}\right)^3
({5 \over 42} a^3 - {33 \over 70} a^2 + {7 \over 10} a
- {1 \over 2} + {4 \over 35} a^{-{1 \over 2}}
+ {4 \over 105} a^{-{3 \over 2}}) \;\;\;, &(4d) \cr
q_{3}^{c} &= \left({3 \over 2}\right)^3
({1 \over 14} a^3 - {3 \over 14} a^2 + {1 \over 10} a
+ {1 \over 2} - {4 \over 7} a^{-{1 \over 2}}
+ {4 \over 35} a^{-{3 \over 2}}) \;\;\;. &(4e) \cr}
$$
The perturbation potentials have to be constructed by
solving iteratively the following set of $7$ Poisson equations:
$$
\eqalignno{
&\Delta_0 {\cal S}^{(1)} = I({\cal S}_{,i,k}) \;t_i \;\;\;,&(4f) \cr
&\Delta_0 {\cal S}^{(2)} = 2 II({\cal S}^{(1)}_{,i,k})
\;\;\;, &(4g) \cr
&\Delta_0 {\cal S}^{(3a)} =
3 III({\cal S}^{(1)}_{,i,k}) \;\;\;, &(4h) \cr
&\Delta_0 {\cal S}^{(3b)} =
\sum_{a,b,c} \epsilon_{abc}\;
{\partial({\cal S}^{(2)}_{,a},{\cal S}^{(1)}_{,b},X_c)
\over \partial (X_1,X_2,X_3)} \;\;\;, &(4i) \cr
(&\Delta_0 {\vec {\cal S}}^{(3c)})_k =
\epsilon_{pq \lbrack j} {\partial ({\cal S}^{(2)}_{,i \rbrack},
{\cal S}^{(1)}_{,p},X_q) \over \partial(X_1,X_2,X_3)}
\;\;\;. &(4j,k,l) \cr}
$$
($i,j,k=1,2,3$ with cyclic ordering).

\bigskip
We realize the solution by first solving Poisson's equation for $\cal S$
via FFT (Fast Fourier Transform) from
the initial density contrast $\delta$ generated as initial data
for the numerical simulation (Section {\bf 2}).
In a flat background model we have:
$$
\Delta_0 {\cal S} = - {2 \over 3 t_i} \delta \;\;.\eqno(5)
$$
We first note that the first-order part of the mapping (4)
is equivalent to Zel'dovich's approximation, if we put
${\cal S}^{(1)} = {\cal S} t_i$. The only formal difference
is that we start at the initial time on an undeformed grid due to
the assumption $a(t_i)=1$: $\vec F(\vec X,t_i)=\vec X$.
This solution of the first
Poisson equation (4f) as well as the entire solution (4) is unique
provided we impose
periodic boundary conditions on $\cal S$ and fix some gauge conditions
accordingly (Ehlers \& Buchert 1993).
With this setting the source expressions in (4f-h) are computable
in terms of the initial data $\cal S$. We solve all derivatives
and all Poisson
equations in $k-$space by using FFT, the four Poisson equations
(4i-l) have to be solved iteratively by using the second-order
perturbation potential calculated first from (4g).

\medskip
The first-order part of the solution (4) covers the kinematical
aspect of the collapse process, whereas the second-order part
covers essential aspects of the tidal action of the gravitational
field (Buchert \& Ehlers 1993).
At the third order, there are also interaction terms
among the perturbation potentials at first and second order, which
induce vorticity in Lagrangian space. Although the solution (4)
is rotational in Lagrangian space, the Eulerian fields remain
irrotational until the first shell-crossings.
The development of caustics will result in multi-stream flow and will
introduce vorticity also in Eulerian space (see, e.g., Chernin 1993).
Note that the
Eulerian representation of the equations breaks down at caustics, while
the Lagrangian representation of the fluid's motion in terms of
trajectories is still regular and formally allows to follow the motion
of the fluid across caustics which we shall do. However, we emphasize
that the Lagrangian solution $\vec f$ represents solutions of the
Euler-Poisson system only as long as the mapping to Eulerian space is
single-valued; In multi-stream regions a formal extrapolation of $\vec
f$ across caustics neglects the gravitational interaction of the
streams
since, in these regions, the gravitational field-strength acting on
{\it each} particle is the sum of the field-strengths of the streams
(compare Ehlers \& Buchert 1993).

\bigskip
In order to understand that the higher-order approximation schemes
are better than the first-order scheme until shell-crossing
(regardless of the initial
conditions and the scale of the fluctuations), the following
general equation can be used:
Consider the
density contrast $\Delta: = (\varrho - \varrho_H) / \varrho$,
$- \infty < \Delta < 1$,
which is more adapted to the non-linear situation than the conventional
definition $\delta=(\varrho - \varrho_H) / \varrho_H = \Delta /
(1 - \Delta)$, defined in Eulerian perturbation theory.
For this the following fully non-linear
evolution equation
derived from the continuity equation and Poisson's
equation has been found (Buchert 1992; see also Peebles 1987, Buchert
1989, Bertschinger \& Jain 1993):
$$\eqalignno{
&\dot \Delta = (\Delta-1) {\bf I} \;\;, &(6a) \cr
&\ddot \Delta + 2 {\dot a \over a} \dot \Delta -
4 \pi G \varrho_H \Delta =
(\Delta-1) 2 {\bf II} \;\;,  &(6b) \cr}
$$
where here ${\bf I}$ and ${\bf II}$
denote the first and second principal scalar invariants, respectively,
of the peculiar-velocity tensor gradient $(u_{i,j})$
(contrary to the notation before, a comma here denotes derivative
with respect to comoving Eulerian coordinates $\vec q$, a dot denotes
Lagrangian time-derivative ${d\over dt}:={\partial \over \partial_t}
\vert_q + {\vec u \over a}\cdot \nabla_q$):
$$
{\bf I}:={1 \over a} \sum_i u_{i,i} \;\;\;,\;\;\;
{\bf II}:={1 \over 2 a^2} \left( (\sum_i u_{i,i})^2 -
\sum_{ij} \left(u_{i,j}u_{j,i}\right)\right)\;\;. \eqno (6c)
$$
This equation is exact. For the approximation schemes we can read
off the following:
The first-order model solves equation (6) with
{\bf II} $\rightarrow 0$,
whereas the second-order model takes the second
invariant into account for small displacements from the first-order
trajectories. Thus, the second-order theory improves upon
first-order until equation (6) breaks down at shell-crossing.

\bigskip
We realize the solution (4) at each order by calculating
$128^3$ trajectories. Note that, although we have $7$ Poisson equations
to solve, the model (4) is very time-efficient: On a CONVEX C220
the third-order model takes about $20$ minutes to realize the
density distribution for each stage
at this resolution, while the first-order model takes $2$ and the
second-order model $3$ minutes. On a CRAY YMP/464 the corresponding
CPU times are smaller by a factor of $5$.
The N-body simulation run here
takes $138$ integration steps each of which uses about 2 minutes on a
CRAY 2, i.e. in total 5.4 hours to get to the final stage.
The N-body code
is limited by the speed of the (rather poor) FFT on a CRAY 2.  Since
this was
written, we have moved our N-body computing to a CONVEX C3.

A higher resolution can be easily obtained for the analytical model,
provided the
smallest wavelength in the model (corresponding to $k_c=4 k_f$ here) is
calculated to sufficient accuracy. For example, if we resolve the box
with length $L$ with
$128$ calculated density values along one spatial direction, the
smallest wavelength
$L/4$ will be calculated at $32$ points; $32$ points will be
sufficient to map the coherence scale at high accuracy.
Interpolating the calculated grid points, e.g.,
$16$ times, we obtain an effective resolution in the
box of $2048^3$ particles. This interpolation method gives highly
accurate results for the final distribution, since the analytical
mapping contains all non-linear information; it maps the interpolated
grid points as if it were calculated ones.
For a detailed discussion and illustration of this method see
(Buchert \& Bartelmann 1991).
\bigskip
Figures 1a to 1d show a comparison
of slices of the N-body model and the first-
through third-order approximate schemes by visual appearance.
Often very high-order
information which is difficult to quantify can be rapidly gained from
visual presentations.

\vfill\eject

\noindent{\rmm 4. Cross-correlation statistics}
\bigskip\medskip
\noindent
{\rma 4.1. Cross-correlation coefficient}
\bigskip\noindent
As in Coles \etal (1993) and Melott \etal (1993)
we use here the usual cross-correlation coefficient $S$ to compare
the resulting density fields:
$$
S := {<(\delta_1 \delta_2)> \over \sigma_1 \sigma_2} \;\;, \eqno(7)
$$
where $\delta_{\ell}, \ell=1,2$  represent the density contrasts in
the analytical and the numerical approximations, respectively,
$\sigma_{\ell} = \sqrt{<\delta_{\ell}^2>-<\delta_{\ell}>^2}$
is the standard deviation in a Gaussian random field;
averages $<...>$ are taken over the entire distribution.
\medskip
The density in the
analytical models is calculated by
collecting trajectories of the Lagrangian perturbation
solutions at the different orders into a $128^3$ pixel grid
with the same mesh method (CIC binning) as in the
N-body simulation.
\medskip
The correlation coefficient obeys the bound $\vert S\vert \le 1$;
$S = 1$ implies that $\delta_1 = {\cal C} \delta_2$, with $\cal C$
constant; ${\cal C}$ close to $1$ implies, of course, better
agreement between the approximation and the numerical simulation.
\medskip
Following Coles \etal (1993) we will compare the approximation with the
simulation with varying amounts of Gaussian smoothing
of each so that we can
follow accuracy at various lengthscales, and allow for
the possibility of
results which are essentially correct but for small
displacements. We will plot
$S$ as a function of $\sigma_2$,
the r.m.s. density contrast in the numerical
simulation.
The cross-correlation coefficient $S$ is calculated down to the
resolution scale of a $128^3$ pixellization of the density field.
\medskip
There will be basic differences from the procedure of Coles \etal.
No truncation
of the spectrum will be considered here because
in pancake models the initial
conditions are already truncated. We will not be
able to compare these results
with those since we are dealing with a different
class of models. Instead, we
will compare the accuracy of first--, second--, and third--order
Lagrangian
approximations with the simulation, and thereby gain information on the
benefits of going to higher--order Lagrangian solutions within
the context of
pancake models. In paper II
on hierarchical models we will determine to what
extent these results carry over to hierarchical models, and we will find
out
whether substantial improvements over the optimally
truncated first-order approach (Melott \etal 1993) is made.
\medskip
In order to couple our cross--correlations to the physical scale,
we show the actual r.m.s. density fluctuations for all approximations
$\sigma_{\ell}$ as a function of the Gaussian smoothing scale in grid
units in
Figure 2 for each of our stages. The Gaussian smoothing is
done by convolution
with $e^{-R^2/2R^2_G}$.
The stages correspond to (Eulerian) linear theory $\sigma_2=0.25,
0.5, 1.0, 1.2, 1.5$, and 2.0 as discussed before.
$R_G$ is given in grid
units; remember that the box is 128 grid units in size, and the smallest
wavelength perturbation in the initial data was 32
grid units. Since it is
known that these Lagrangian schemes give rather accurate
locations for
pancakes, we are more interested in studying details of structure.
Therefore most
of our information will be relevant for $R_G<32$ grid units.
\medskip
In Figures 3a to 3f we show our most important basic result, the
cross-correlation coefficient
of the various orders of approximation with the N-body simulation at
various stages.

\bigskip\medskip\noindent
{\rma 4.2. Power spectrum analysis}
\bigskip\noindent
Since Lagrangian perturbation solutions include substantial
non-linearities (see, e.g., Buchert 1992), we expect that they
do not preserve the initial power spectrum
of the fluctuation field as the Eulerian linear solution
would do. Therefore, it has meaning to test the agreement with the
N-body simulation.
For the spectrum considered in the present paper ($n=-1$) Melott \etal
(1993) found that the power spectrum of the
truncated ZA with that of the N-body
simulation is substantially underestimated. In this respect
the spectrum with index $n=-1$ provides an extreme case. We,
here, analyze whether the higher-order corrections can improve on that,
and present the results of the power spectrum analysis in Figures 4a to
4f.

\bigskip\medskip\noindent
{\rma 4.3. Relative phase-errors}
\bigskip\noindent
The non-linear evolution of the fluctuation field does not preserve
the initial power spectrum as well as the phase information, which is
not
contained in the power spectrum. There will be a substantial change in
the phases due to non-linear phase-locking effects, which are
already described by the first-order Lagrangian perturbation solutions
(see Melott \etal 1993). Also here we expect an improvement due to
higher-order corrections, since they contain essential effects of tidal
interactions which should contribute to the change of the phases.
\medskip\noindent
We calculate the {\it relative phase error} as follows:
The Fourier coefficients of the initial fluctuation field
$\delta_{\vec k} = \vert \delta_{\vec k}\vert e^{i\phi_{\vec k}}$
contain
information about the amplitude $\vert \delta_{\vec k}\vert$ and the
phase angle $\phi_{\vec k}$.
We measure the angle $\theta = \phi_1 - \phi_2$,
where $\phi_2$ are the phases in the N-body simulation, and $\phi_1$
in the analytical approximations.
We present the results on the relative phase errors in terms of
$\cos(\theta)$ as a function of $k$ (calculated in spherical shells in
$k-$space).
The opposite assignment $\theta = \phi_2 - \phi_1$ yields equivalent
results since $\cos(\theta)$ is symmetric around $\theta=0$.
Perfect agreement between the N-body result
and the analytical scheme implies $\cos(\theta)=1$, anti-correlated
phases have $\cos(\theta)=-1$, and for randomized phases $\cos(\theta)$
would average to $0$.
We present the result in Figures 5a to 5f.

\bigskip\medskip\noindent
{\rma 4.4. Density distribution functions}
\bigskip\noindent
The cross-correlation coefficient ({\bf 4.1}) measures differences
in position up to a constant ratio ${\cal C}$ of the compared densities
(compare Section {\bf 2}).
In order to probe the differences of the various approximations
with respect to the actually predicted densities on different spatial
scales, we calculate the density distribution function for each
approximation and depict them in
Figures 6a to 6d in terms of the number of cells $N$ found with
mass density $\rho$ (in units of the mean) as a function of scale.

\bigskip\bigskip\bigskip

\noindent{\rmm 5. Discussion of the results}
\bigskip\medskip
\noindent
In Figures 1a to 1d we show slices of the density fields corresponding
to the stages $2,3,4$ and
$5$, while the statistical analysis is done for all six stages.
\medskip
At the first stage ($\sigma_2=0.25$) pancake formation is just
beginning.
Here, the higher-order schemes predict structures which are almost
identical to the N-body result, whereas
the first-order scheme appears to delay the collapse
time as was expected. The cross-correlation coefficient as well as the
power spectrum measures this feature clearly (Figs. 3a, 4a); the
phase information is
represented better at higher orders especially on small scales (Fig.
5a). Also the density distribution functions confirm this behavior
(Fig. 6a).
The number of high-density cells is underestimated by all
schemes.
\medskip\noindent
Stage $2$ ($\sigma_2=0.5$, Figs. 1a) emphasizes this: Here, the
higher-order schemes significantly improve on the first-order
``Zel'dovich approximation''; the cross-correlation coefficient (Fig.
3b)
is as high as $S=0.90$ for the second-order and only slightly higher for
the third-order scheme, while
the first-order coefficient has $S=0.85$ on the scale where
$\sigma_2=1$.
A similar conclusion can be drawn from
the power spectrum analysis (Fig. 4b). However, the power on small
scales is already underestimated by all schemes, but represented
better at higher orders.
Interestingly, the phase information contained in the first-order
approximation
shows randomized relative phase-errors, while the higher-order schemes
show a strong tendency to anti-correlated phases at this stage
(Fig. 5b). This point would become
clearer by analyzing the velocity fields.
The better agreement with respect to the cross-correlation and the
power spectrum is also a consequence of a
better prediction of the collapse time in the higher-order solutions.
The density distribution functions (Fig. 6b) show good agreement
with the N-body distribution for the second- and third-order models.
The first-order model still suffers from its properties mentioned
before at stage $1$.
\medskip\noindent
At stage $3$ (corresponding to the present stage according to the
standard normalization, $\sigma_2=1$) we can already see slight
differences in the density fields (Figs. 1b): The biggest difference
is apparent in the lower-right corner where the most evolved structure
contained in this slice is seen. We appreciate that in the higher-order
solutions filaments stay more
compact like in the N-body slice in contrast to
the first-order solution.
The cross-correlation coefficient (Fig. 3c) shows an almost
constant improvement of the second-order upon the first-order
approximation down to the smallest scales,
while the third-order scheme becomes worse than the
second-order result on scales $\sigma_2>2$, but still is
better than that of first-order up to $\sigma_2 \approx 4$.
The power spectrum (Fig. 4c) remains better represented in the
higher-order solutions.
Starting from this stage, the relative phase-errors show a tendency
to randomization for all orders of the perturbation solutions, while
the scale where randomization occurs increases with time
(Figs.
5c-f). The density distribution functions (Fig. 6c) show excellent
agreement with the numerical result but begin to overestimate the
number of cells with moderately high density, a feature which
is due to the overestimated size of the pancakes.

\medskip\noindent
Stage $4$ ($\sigma_2=1.2$) is the most interesting stage, where all
orders of
approximation
schemes provide equally good results. The agreement of the
density
fields with the N-body simulation is still reasonable in all schemes.
The most evolved structure shows compact internal density peaks inside
the first-order pancakes
(Figs. 1c). The cross-correlation coefficients (Fig. 3d)
are still high, but only slightly higher in the second- and third-order
 approximations
on large scales corresponding to
$\sigma_2 < 3$ (second-order), $\sigma_2
< 2.5$ (third-order). On smaller scales the
second-order
coefficient stays close to the first-order coefficient, while the
third-order scheme starts to fall off more rapidly below the
cross-correlation of the first-order scheme.
The power spectra (Fig. 4d) are almost identical for all schemes,
a property which roughly remains in the following stages.
Here, the density distribution functions (Fig. 6d) mirror an
amplification of
the effect mentioned for stage $3$.

\medskip\noindent
The next stage $5$ ($\sigma_2=1.5$) we can identify with the stage
where the
break-down of the Lagrangian perturbation solutions has started.
Figs. 1d show differences in the thickness of pancakes,
which begin to grow more rapidly in the perturbation solutions
than in the N-body simulation.
The density distribution functions (Fig. 6e) overestimate the
number of low-density cells, while the number of high-density cells
is underestimated.
The scale above which an improvement of the higher-order
schemes can be detected shifts to larger scales, here the scale
is roughly $\sigma_2 \approx 1.5$ (Fig. 3e). Furthermore, the
improvement is
now very small. The power spectrum of higher order schemes is now
than
worse than first--order on small scales (Fig. 4e).
\medskip\noindent
At stage $6$ ($\sigma_2=2.0$) such a scale no longer exists, and we
cannot appreciate any improvement of the higher-order schemes upon
first-order. The break-down of the perturbation sequence is reached. To
summarize, stage 3 $(\sigma_2=1.0)$ is the last stage at which the
improvement
of second-- and third--order schemes over first is large and positive.
Increasingly in stage 4 and later differences become insignificant.
There is
substantial benefit
at early stages from second--order, but there is never any
important additional improvement by going to the
third--order.

\bigskip\bigskip\bigskip

\noindent{\rmm 6. Conclusions}
\bigskip\medskip
\noindent
We have analyzed a time sequence of stages in a pancake model simulated
numerically and compared with analytical Lagrangian perturbation
solutions at various orders. We found that
until the stage when $\sigma_2$(linear theory)$=1$
(corresponding to the present epoch according to the standard
normalization of large-scale structure models), the higher-order
Lagrangian solutions clearly improve upon the first-order
``Zel'dovich approximation'' down to the resolution scale of the
simulation. At later stages any small improvement can only be
appreciated
above a scale which increases with time until the stage
$\sigma_2$(linear theory)$\approx 1.5$.
Later stages correspond to the onset of hierarchical
clustering, which are beyond the reach of the perturbation solutions
at any order in this model. Also, third--order is computationally
different and
does not improve things much beyond second--order at any stage.

A typical feature of this break-down is the behavior
of the higher-order schemes, which both become worse than the
first-order
scheme at later stages. This is due to the fact that in a perturbation
approach the
higher-order time coefficients grow more rapidly than the first-order
time coefficient (here: $q_1 \propto a$, $q_2 \propto a^2$, $q_3 \propto
a^3$). Consequently, the higher-order corrections blow up at and after
the time coefficient functions are of the same order in all schemes
which is the case roughly at the stage $\sigma_2$(linear theory)$=1.5$.
The limit (or break-down) scale $k_b$ above which an
improvement of the higher-order schemes upon the
``Zel'dovich approximation'' can be detected at about
$\sigma_2(k_b)\approx 2$ corresponds to $R_G \approx 5$
and shifts to larger scales at later stages.
After this stage we probe the onset of hierarchical clustering,
for which the perturbation solutions are not meant for in the
first place. Whether a truncation of the power spectrum in
hierarchical models can retain the improvements detected here, will be
analyzed in paper II. We also think that the break-down of the
approximations must not be attributed to the lack of higher-order
corrections. It might well be that the neglection of the
self-gravitating interaction of streams in multi-stream regions by
simply extrapolating the trajectories across caustics must be
considered a major source of this break-down phenomenon.

We emphasize that both the N-body simulation and the Lagrangian
perturbation solutions are assumed to be approximations to the unknown
exact solution.

Another result of this paper is the relative importance of terms in the
higher-order schemes, if we are concerned with the large-scale
performance of the solutions in pancake models: We found that the
vector perturbation potential ${\cal S}^{(3c)}$ can be safely neglected
in this case. It will only be relevant if we resolve internal
structures of pancakes. The same can be said for
the third-order
correction as a whole which does not improve upon the
second-order scheme on large scales except with respect to the
density distribution function at early stages.
However, internal structures of
pancakes are affected due to the prediction of a third shell-crossing
inside pancakes; also the collapse occurs substantially earlier than
in the second-order model if calculated at high resolution
(see Buchert \etal 1993). The higher core density in
clusters as a result of
this does not yield a noticeable improvement in the cross-correlation
with the N-body simulation in the present case except a slight
improvement of the cross-correlation of the third-order model on small
scales at later stages.
We also found that the realization of the third-order model is
extremely sensitive to numerical errors by iteratively solving
Poisson equations. A direct analytical calculation by explicitly
solving the Poisson equations for a given model
is more reliable but extremely CPU time extensive for a generic
fluctuation field (see: Buchert \etal 1993).
The second-order scheme which is much easier and faster to realize
provides the main effect of improvement upon the ``Zel'dovich
approximation''. It also provides a good compromise between improvement
at early stages, while the surprisingly good performance of the
first-order scheme on small scales at later stages is essentially
preserved.
Here we emphasize that the first impression of moderate
improvement of second upon first order is misleading, since
we compare with an approximation which already shows excellent
agreement with the N-body result: e.g., the cross-correlation
coefficient is already high in the first-order scheme.
The latest stages analyzed here are in a regime where we are moving into
hierarchical clustering.  A removal of
unwanted non-linearities by using, e.g., a Gaussian smoothing of
the initial data
will definitively change this picture (Melott \etal 1993).
We will examine this when
we study hierarchical models in paper II.
We finally note that second-order solutions are available
for a larger class of initial data than considered in this paper
(Buchert \& Ehlers 1993).

To summarize, without reservation we recommend the use of second--order
perturbation theory up to the stage $\sigma$(linear theory)$=1$,
as a definite and useful
improvement upon first--order
(the ``Zeldovich approximation'') in pancake models.

\bigskip
The numerical code to realize Lagrangian perturbation solutions
is available via e-mail: TOB @ ibma.ipp-garching.mpg.de $\;\;$.

\bigskip\bigskip\noindent
{\rma Acknowledgements:} {\mfst We would like to thank Robert Klaffl
for help during the preparation stage of this work and for valuable
discussions as well as J\"urgen Ehlers for valuable remarks on the
manuscript.

\noindent
TB is supported by DFG (Deutsche Forschungsgemeinschaft). ALM wishes to
acknowledge support from NASA Grant NAGW--2923, NSF grants AST--9021414
and
OSR--9255223, and facilities of the National Center for Supercomputing
Applications, all in the USA. }

\vfill\eject

\def\ref{\par\noindent\hangindent\parindent\hangafter1}
\centerline{\rmm References}
\bigskip\bigskip
{\mfst
\ref
Beacom J.F., Dominik K.G., Melott A.L., Perkins S.P., Shandarin S.F.
(1991): {\it Ap.J.} {\bf 372}, 351.
\smallskip
\ref
Bertschinger E., Jain B. (1993): {\it Ap.J.}, in press.
\smallskip
\ref
Blanchard A., Buchert T., Klaffl R. (1993): {\it Astron. Astrophys.}
{\bf 267}, 1.
\smallskip
\ref
Bouchet F.R., Juszkiewicz R., Colombi S., Pellat R. (1992):
{\it Ap.J. Lett.} {\bf 394}, L5.
\smallskip
\ref
Buchert T. (1989): {\it Astron. Astrophys.} {\bf 223}, 9.
\smallskip
\ref
Buchert T., Bartelmann M. (1991): {\it Astron. Astrophys.} {\bf 251},
389.
\smallskip
\ref
Buchert T. (1992): {\it M.N.R.A.S.} {\bf 254}, 729.
\smallskip
\ref
Buchert T. (1993a): {\it Astron. Astrophys. Lett.} {\bf 267}, L51.
\smallskip
\ref
Buchert T. (1993b): {\it M.N.R.A.S.}, in press.
\smallskip
\ref
Buchert T., Ehlers J. (1993): {\it M.N.R.A.S.} {\bf 264}, 375.
\smallskip
\ref
Buchert T., Karakatsanis G., Klaffl R., Schiller P. (1993):
{\it Astron. Astrophys.}, to be submitted.
\smallskip
\ref
Chernin A.D. (1993): {\it Astron. Astrophys.} {\bf 267}, 315.
\smallskip
\ref
Coles P., Melott A.L., Shandarin S.F. (1993): {\it M.N.R.A.S.}
{\bf 260}, 765.
\smallskip
\ref
Croudace K.M., Parry J., Salopek D.S., Stewart J.M. (1993): {\it Ap.J.},
in press.
\smallskip
\ref
Ehlers J., Buchert T. (1993): in preparation.
\smallskip
\ref
Giavalisco M., Mancinelli B., Mancinelli P.J., Yahil A. (1993):
{\it Ap.J.} {\bf 411}, 9.
\smallskip
\ref
Gramann M. (1993): {\it Ap.J. Lett.} {\bf 405}, 47.
\smallskip
\ref
Kofman L.A., Pogosyan D., Shandarin S.F., Melott A.L. (1992):
{\it Ap.J.} {\bf 393}, 437.
\smallskip
\ref
Lachi\`eze-Rey M. (1993): {\it Ap.J.} {\bf 408}, 403.
\smallskip
\ref
Matarrese S., Lucchin F., Moscardini L., Saez V. (1992): {\it
M.N.R.A.S.} {\bf 259}, 437.
\smallskip
\ref
Melott A.L. (1984): {\it Sov. Astron.} {\bf 28}, 631.
\smallskip
\ref
Melott A.L. (1986): {\it Phys. Rev. Lett.} {\bf 56}, 1992.
\smallskip
\ref
Melott A.L. (1987): {\it M.N.R.A.S.} {\bf 228}, 1001.
\smallskip
\ref
Melott A.L., Shandarin S.F. (1990): {\it Nature} {\bf 346}, 633.
\smallskip
\ref
Melott A.L., Pellman T.F., Shandarin S.F. (1993): {\it M.N.R.A.S.},
submitted.
\smallskip
\ref
Melott A.L., Shandarin S.F. (1993): {\it Ap.J.} {\bf 410}, 469.
\smallskip
\ref
Melott A.L. (1993): {\it Ap.J. Lett.} {\bf 414}, L73.
\smallskip
\ref
Moutarde F., Alimi J.-M., Bouchet F.R., Pellat R., Ramani A.
(1991): {\it Ap.J.} {\bf 382}, 377.
\smallskip
\ref
Peebles P.J.E. (1980): {\it The Large-scale Structure of the Universe},
Princeton Univ. Press.
\smallskip
\ref
Peebles P.J.E. (1987): {\it Ap.J.} {\bf 317}, 576.
\smallskip
\ref
Shandarin S.F., Zel'dovich Ya.B. (1984): {\it Phys. Rev. Lett.} {\bf
52}, 1488.
\smallskip
\ref
Shandarin S.F., Zel'dovich Ya.B. (1989): {\it Rev. Mod. Phys.} {\bf 61},
185.
\smallskip
\ref
Weinberg D.H., \etal (1993): in preparation.
\smallskip
\ref
Zel'dovich Ya.B. (1970): {\it Astron. Astrophys.} {\bf 5}, 84.
\smallskip
\ref
Zel'dovich Ya.B. (1973): {\it Astrophysics} {\bf 6}, 164.

}

\bigskip\bigskip\bigskip

\centerline{\rmm Figure Captions}
\bigskip\medskip
{\mfst
\noindent{\bf Figure 1:} A thin slice (thickness $L/128$) of the
density field is shown for the numerical (upper left), the first-order
(upper-right), the second-order
(lower left), and the third-order (lower right) approximations for
the evolution stages 2 ($\sigma_2 = 0.5$, Fig. 1a),
3 ($\sigma_2 = 1.0$, Fig. 1b),
4 ($\sigma_2 = 1.2$, Fig. 1c), and 5 ($\sigma_2 = 1.5$, Fig. 1d).
The grey-scale is linear, and the maximum density contrasts
are chosen to be $8$ in Fig. 1a, and $25$ in Figs.
1b,c,d. (The slight stripes in low-density regions are artifacts
resulting from the interaction of the distorted grid of the particles
with the pixelization.)

\bigskip

\noindent{\bf Figure 2:} The standard deviations of the density
contrast as a function of smoothing scale $R_G$
in the first-order (dotted),
the second-order (dashed), the third-order (dashed-dotted), and the
numerical (full line) approximations, for the
evolution stages 1,2,3,4,5,6 (Figs.
2a,b,c,d,e,f). $R_G$ is given in grid units; the shortest wave in the
initial data was 32 grid units.

\bigskip

\noindent{\bf Figure 3:} The cross-correlation coefficient $S$
as a function of the standard deviation $\sigma_2$ for the
evolution stages 1,2,3,4,5,6 (Figs. 3a,b,c,d,e,f).
The cross--correlation of the N-body
with first-order perturbation theory is shown as a dotted line;
with second-order
a dashed line; and with third-order a dashed-dotted line.

\bigskip

\noindent{\bf Figure 4:} The power spectra of the N-body simulation
(solid line) compared with the first- (dotted), second- (dashed),
and third-order (dashed-dotted) approximation schemes for the stages
$1,2,3,4,5,6$ (Figs. 4a,b,c,d,e,f).

\bigskip

\noindent{\bf Figure 5:} The relative phase-errors. Notation and
labelling like in Figure 3.

\bigskip

\noindent{\bf Figure 6:} The density distribution functions. Notation
and labelling like in Figure 4.

}

\vfill\eject
\bye